\documentclass[12pt]{article}
\usepackage{amsmath,amssymb}
\usepackage{cite}
\numberwithin{equation}{section}

\newcommand{\ba}{\begin{eqnarray}}
\newcommand{\ea}{\end{eqnarray}}

\newcommand{\deltam}[2]{\delta^{{\rm mod}\  #2}_{#1}}
\newcommand{\Th}{\Theta}
\newcommand{\Ncal}{{\cal N}}

\newcommand{\nn}{\nonumber}
\newcommand{\Zb}{\mathbb{Z}}
\newcommand{\Rb}{\mathbb{R}}

\newcommand{\e}[1]{{\bf e}\!\left[#1\right]}

\DeclareMathOperator{\lcm}{{\rm lcm}}

\newcommand{\slr}{SL(2,\mbox{$\Rb$})}
\newcommand{\chic}{\check{\chi}}
\newcommand{\cc}{\check{c}}
\newcommand{\basi}{{\bf bas}}
\newcommand{\vect}{{\bf vec}}
\newcommand{\spi}{{\bf spi}}
\newcommand{\cospi}{{\bf cos}}

\newcommand{\so}{{\rm so}}

\newcommand{\su}{{\rm su}}
\newcommand{\SU}{\su}

\newcommand{\isi}{{\bf Ising}}
\newcommand{\tri}{{\bf Tri}}
\newcommand{\pot}{\text{\bf 3-Potts}}

\newcommand{\Ghol}{G_{\rm hol}}

\begin{document}

\thispagestyle{empty}
\begin{flushright}
 \parbox{3.5cm}{KUCP-199\\ {\tt hep-th/0112004}}
\end{flushright}

\vspace*{1cm}
\begin{center}
 {\Large
    Coset Character Identities in Superstring Compactifications
 }
\end{center}

\vspace*{2cm} 
\begin{center}
 \noindent
 {\large Satoshi Yamaguchi}

 \vspace{5mm}
 \noindent
 \hspace{0.7cm} \parbox{142mm}{\it
 Graduate School of Human and Environmental Studies,
 Kyoto University, Yoshida-Nihonmatsu-cho,
 Sakyo-ku, Kyoto 606-8501, Japan.

 E-mail: {\tt yamaguch@phys.h.kyoto-u.ac.jp}
 }
\end{center}

\vspace{4cm}
\hfill{\bf Abstract\ \ }\hfill\ \\
We apply the coset character identities (generalization of Jacobi's
abstruse identity) to compact and noncompact Gepner models. In the both
cases, we prove that the partition function actually vanishes due to the
spacetime supersymmetry. In the case of the compact models and discrete
parts of the noncompact models, the partition function includes the
expected vanishing factor.  But the character identities used to the
continuous part of the noncompact models suggest that these models have
twice as many supersymmetry as expected. This fact is an evidence for
the conjecture that the holographically dual of the string theory on an
actually singular Calabi-Yau manifold is a super {\em conformal} field
theory.  The extra SUSY charges are interpreted as the superconformal S
generators.

\newpage

\section{Introduction}
The supersymmetric compactification of the string theory is an
important problem in both the phenomenological and the formal
sense. The internal space of the compactification should be a
manifold of a special holonomy if we want to preserve some of the
spacetime supersymmetry.

From the point of view of the worldsheet conformal field theory, the
spacetime supersymmetry implies that the toroidal partition function
should vanish. For example, if we consider the string theory on the flat
spacetime, toroidal partition function vanishes because of the Jacobi's
abstruse identity: the relation among Jacobi's theta functions
\begin{align*}
 \theta_3(\tau)^4-\theta_4(\tau)^4-\theta_2(\tau)^4=0.
\end{align*}
Let us use the affine \so(8) characters $\chi^{\so(8)}_{s}$.  The
subscript ``$s$''$(=\basi,\vect,\spi,\cospi)$ respectively expresses the
basic, spinor, vector, and cospinor representation of \so(8). Then the
Jacobi's abstruse identity can be written as
\begin{align*}
 \chi^{\so(8)}_{\vect}(\tau)=\chi^{\so(8)}_{\spi}(\tau).
\end{align*}
This equation shows that the number of the bosons (in the vector
representation of affine \so(8)) is equal to the number of the fermions
(in the spinor representation of affine \so(8)) in each mass level.
This is the result of the spacetime supersymmetry.

Then, how about the compactification with a manifold of nontrivial
special holonomy $\Ghol$? In \cite{Sugiyama:2001qh}, it has been
proposed that the coset CFT \so(8)$/\Ghol$ is essential to the spacetime
supersymmetry. Especially, the character identity used to show that the
partition function vanishes is
\begin{align}
 \chi^{\so(8)/\Ghol}_{\vect,\lambda}(\tau)
      =\chi^{\so(8)/\Ghol}_{\spi,\lambda}(\tau),
 \label{susyIdentity}
\end{align}
where $\lambda$ is a representation of level 1 affine $\Ghol$ and
$\chi^{\so(8)/\Ghol}_{s,\lambda}(\tau)$ is a coset character.  Each
character identity of the case $\Ghol=\SU(2),\SU(3),G_2$ has been found
previously in \cite{Bilal:1987uh, Eguchi:1989vr,%
Kutasov:1991pv,Eguchi:2001xa}. Many other character identities related
to SU(2)/U(1) coset models have been found in \cite{Argyres:1993fr}.

In this paper, we consider the noncompact Gepner models
\cite{Mizoguchi:2000kk,Eguchi:2000tc,Yamaguchi:2000dz,Mizoguchi:2000sz,Naka:2000uy,Yamaguchi:2001rn}.
The partition functions of the models in
\cite{Mizoguchi:2000kk,Eguchi:2000tc,Mizoguchi:2000sz,Naka:2000uy} have
been already shown to vanish in each paper.  But, the models in
\cite{Yamaguchi:2000dz,Yamaguchi:2001rn} have not yet been shown to have
vanishing partition functions. In this paper, we show these partition
functions actually vanish because of the character identity
(\ref{susyIdentity}).

Among these models, the continuous part of the noncompact Gepner models
includes larger coset models than expected from their holonomies, as we
show in this paper. That is, the partition function of the noncompact
Gepner models of an ALE compactification includes the affine $\so(8)$
itself, the partition function of a Calabi-Yau 3-fold compactification
includes the coset $\so(8)/\su(2)$, and the partition function of a
Calabi-Yau 4-fold compactification includes the coset
$\so(8)/\su(3)$. This fact is an evidence for the conjecture that
holographically dual field theory of the string theory on an actually
singular Calabi-Yau manifold is a super {\em conformal} field
theory\cite{Gukov:1999ya,Giveon:1999zm}.  Generally, a superconformal
theory has the superconformal S generators besides the ordinary Q
generators.  The extra spacetime supercharges in the singular Calabi-Yau
models correspond to the superconformal S generators of the dual
superconformal field theory.

The organization of this paper is as follows.
In section \ref{characterIdentity}, we review the character identities.
In section \ref{compact}, we treat the compact
Gepner models. We show that a compact Gepner model includes the
appropriate coset models. In section \ref{noncompact}, we treat the
noncompact Gepner models. The last section \ref{conclusion} is
devoted to conclusions and discussions.

\section{Character identities} \label{characterIdentity}
In this section, we review the character identities of
eq.(\ref{susyIdentity}) and their properties. 

The coset CFT $\so(8)/\Ghol$ is essential to the spacetime supersymmetry
of a superstring compactification with a manifold of special holonomy
$\Ghol$. The character of this coset CFT is defined by
the branching relation
\begin{align}
 & \chi^{\so(8)}_{s}=
     \sum_{\lambda}\chi^{\so(8)/\Ghol}_{s,\lambda}
      \chi^{\Ghol}_{\lambda}.
\label{branchingRelation}
\end{align}
In order to express the character identities shortly, we define the
vanishing functions $\xi^{\Ghol}_{\lambda}$ as
\begin{align*}
 \xi^{\Ghol}_{\lambda}=\chi^{\so(8)/\Ghol}_{\vect,\lambda}
             -\chi^{\so(8)/\Ghol}_{\spi,\lambda}.
\end{align*}
This function $\xi^{\Ghol}_{\lambda}$ express the difference between the
number of bosons and the number of fermions.  By using this function
$\xi^{\Ghol}_{\lambda}$, the character identity (\ref{susyIdentity}) can
be written as $\xi^{\Ghol}_{\lambda}=0$ . In order to consider the flat
case as in the same manner, it is useful to denote
$\xi^{\{1\}}:=\chi^{\so(8)}_{\vect}
-\chi^{\so(8)}_{\spi}$. The Jacobi's abstruse identity can be
written as $\xi^{\{1\}}=0$. With these notations and the branching
relation (\ref{branchingRelation}), we obtain the equation
\begin{align}
 & \xi^{\{1\}}=\sum_{\lambda}\xi^{\Ghol}_{\lambda}
      \chi^{\Ghol}_{\lambda}.
 \label{branchingRelation2}
\end{align}
The left hand side of this equation vanishes due to the Jacobi's
abstruse identity and this is an evidence for the character identities
$\xi^{\Ghol}_{\lambda}=0$ .

The modular transformation laws of these functions are
necessary in order to construct modular invariant partition functions.
 The modular properties of $\xi^{\{1\}}$
 become
\begin{align*}
 &\xi^{\{1\}}(\tau+1)=\e{\frac13}\xi^{\{1\}}(\tau),
 &\xi^{\{1\}}(-1/\tau)=\xi^{\{1\}}(\tau),
\end{align*}
where $\e{x}:=\exp(2\pi i x)$. By using these formulae,
the modular properties of
 $\xi^{\Ghol}_{\lambda}$ can be read from the branching relation
(\ref{branchingRelation2}) and they become
\begin{align*}
 &\xi^{\Ghol}_{\lambda}(\tau+1)=\e{\frac 13-h_{\lambda}+c_{\Ghol}/24}
        \xi^{\Ghol}_{\lambda}(\tau),
 &\xi^{\Ghol}_{\lambda}(-1/\tau)=\sum_{\lambda'}
      \bar S^{\Ghol}_{\lambda\lambda'}\xi^{\Ghol}_{\lambda'}(\tau),
\end{align*}
where $h_{\lambda}$ is the conformal dimension of the representation
$\lambda$, $c_{\Ghol}$ is the central charge of the level $1$ affine
$\Ghol$, and $S^{\Ghol}_{\lambda\lambda'}$ is the modular S matrix of
the level $1$ affine $\Ghol$.

Note that the relation (\ref{susyIdentity}) is satisfied not for all
$\Ghol$ and embeddings. It is clear that when
$\chi^{\so(8)}_{\vect,\lambda}\ne0$ and $\chi^{\so(8)}_{\spi,\lambda}=0$
or vice versa from the selection rules, the relation (\ref{susyIdentity})
is not satisfied.

In this paper, we treat the cases $\Ghol=\su(2),\su(3),$ and $G_2$.
In these cases, the relation (\ref{susyIdentity})
is satisfied. Let us see the relation explicitly for each case.

\subsection{\su(2) holonomy case}
The integrable highest weight representations of the level $1$ affine
\su(2) are the basic representation and the fundamental representation.
We denote these two representations by $a=0,1$, respectively.
Then, the vanishing function $\xi^{\su(2)}_{a}$ becomes
\begin{align}
  \xi^{\su(2)}_{a}=\frac{1}{\eta^3}\sum_{s\in \Zb_4}(-1)^s
  \Theta_{s,2}\Theta_{s+2a+1,2}\Theta_{s+a,1} .
  \label{su(2)holonomyidentity}
\end{align}
$\xi^{\su(2)}_{a}=0$ is the theta identity found in \cite{Bilal:1987uh}.

Actually, $\xi^{\su(2)}_{a}$ is related to $\xi^{\su(3)}_{a}$ .
We can prove the identity for the \su(2) holonomy
by using the result of the identity for the \su(3) holonomy.
We mention this proof in the next subsection after we see the
form of the $\xi^{\su(3)}_{a}$.

\subsection{\su(3) holonomy case}

The integrable highest weight representations of the level $1$ affine
\su(3) are the basic, the fundamental, and the conjugate fundamental
representation.  We denote these three representations by $a=0,1,-1$,
respectively.  The explicit form of the $\xi^{\su(3)}_{a}$ can be
written as
\begin{align}
 \xi^{\su(3)}_{a}=\frac{1}{\eta^2}\sum_{s\in\Zb_4}(-1)^{s}\Theta_{6+4a-3s,6}
     \Theta_{s,2} .\label{su(3)holonomyidentity}
\end{align}
The identity $\xi^{\su(3)}_{a}=0$ is found in 
\cite{Eguchi:1989vr,Kutasov:1991pv}.
 
We can prove the character formula $\xi^{\su(3)}_{a}=0$
by explicit calculation. By using the product
formula (\ref{2product-formula})
and $\Theta_{m,k}(\tau)=\Theta_{-m,k}(\tau)$ , $\xi^{\su(3)}_{a}$ can be
written as
\begin{align*}
 \xi^{\su(3)}_{a}&=\frac{1}{\eta^2}\sum_{s\in\Zb_4}(-1)^s\Theta_{6-4a+3s,6}
     \Theta_{s,2}
     &=\frac{1}{\eta^2}
          \sum_{r\in\Zb_8}\Theta_{-12+8a+24r,96}
    \sum_{s\in\Zb_4}(-1)^s\Theta_{-2+4(s+r-a+2),8}.
\end{align*}
In this equation, the second sum vanishes because of the equation
\begin{align*}
 \sum_{s\in\Zb_4}(-1)^s\Theta_{-2+4(s+r-a+2),8}=
(-1)^{r-a}(\Theta_{-2,8}-\Theta_{2,8}+\Theta_{-6,8}-\Theta_{6,8})
 =0.
\end{align*}
We can conclude that the identity $\xi^{\su(3)}_{a}=0$ is actually satisfied.

As mentioned in the previous subsection, we can write the relation
between $\xi^{\su(3)}_{a}(\tau),\ a=-1,0,1$ and 
$\xi^{\su(2)}_{a}(\tau),\ a=0,1$ as
\begin{align*}
 \xi^{\su(2)}_{a}=\sum_{b=-1,0,1}\frac{\Theta_{3a+2b+1,3}}{\eta}
       \xi^{\su(3)}_{b}.
\end{align*}
By using this formula and the proven identity $\xi^{\su(3)}_{a}(\tau)=0$ ,
we can prove the character identity $\xi^{\su(2)}_{a}(\tau)=0$.

The $\xi^{\su(3)}_{a}$ and the $\xi^{G_2}_{a}$ are also related.
We express this relation in the next subsection.
\subsection{$G_2$ holonomy case}
The integrable highest weight representations of the level $1$ affine $G_2$ 
are the basic and the fundamental representation.
We denote these representations by $a=0,1$, respectively.
The explicit form of the $\xi^{G_2}_{a}$ can be written as
\begin{align*}
 &\xi^{G_2}_{0}=\chi^{\isi}_{1/2}\chi^{\tri}_{0}
     +\chi^{\isi}_{0}\chi^{\tri}_{3/2}
    -\chi^{\isi}_{1/16}\chi^{\tri}_{7/16},\\
 &\xi^{G_2}_{1}=\chi^{\isi}_{1/2}\chi^{\tri}_{3/5}
   +\chi^{\isi}_{0}\chi^{\tri}_{1/10}
 -\chi^{\isi}_{1/16}\chi^{\tri}_{3/80},
\end{align*}
where the $\chi^{\isi}$'s and $\chi^{\tri}$'s are the Virasoro
characters of the Ising and tricritical Ising model, respectively.  The
explicit forms of these characters are shown in the Appendix
\ref{appendix-minimal}.  The identity $\xi^{G_2}=0$ is found in
\cite{Eguchi:2001xa}.

As mentioned in the previous subsection, $\xi^{G_2}_{a}$ and
$\xi^{\SU(3)}_{a}$ are related by the equations
\begin{align*}
&\xi^{\SU(3)}_{0}=\xi^{G_2}_{0}C^{\pot}_{0}+\xi^{G_2}_{1}C^{\pot}_{5/2},\\
&\xi^{\SU(3)}_{\pm 1}=\xi^{G_2}_{0}C^{\pot}_{2/3}+\xi^{G_2}_{1}C^{\pot}_{1/15},
\end{align*}
where $C^{\pot}$'s are the $W_3$ characters of the 3-state Potts model.
The explicit forms of these characters are shown in the Appendix
\ref{appendix-minimal}.

\section{Compact Gepner models}\label{compact}
In this section, for a warming up, we show that the partition functions
of the compact Gepner models for Calabi-Yau 3-fold compactifications
include $\xi^{\SU(3)}_{a}$. This fact is shown in \cite{Eguchi:1989vr}
by spectral flow method.  In this paper, we use another method of
construction --- beta method \cite{Gepner:1988qi}.

The Gepner's construction is as follows.  We will work in the lightcone
gauge. Then, the total central charge should be $12$. We consider the
model with 4-dimensional flat spacetime, then we need $c=9$ theory for
the internal space.  The direct product of $\Ncal=2$ minimal models is
used to this internal space.  The total theory in lightcone gauge
includes $R$ of the $\Ncal=2$ minimal models (internal space), and 2
pairs of free bosons and fermions (transverse directions of spacetime)
\begin{align*}
 \Rb^2\times M_{N_1}\times\dots\times M_{N_R},
\end{align*}
where $M_N$ stands for the level $(N-2)$ $\Ncal=2$ minimal model.
Since the total central charge should be $c=12$, the following relation
holds
\begin{align*}
 1+\sum_{j=1}^{R}\frac{N_j-2}{N_j}=4.
\end{align*}

Now, let us construct the modular invariant partition function,
by following \cite{Gepner:1988qi}.
Let us define the characters of the total theory as
\begin{align*}
 \chi^{\lambda}_{\mu}(\tau,z):=\chi^{\so(2)}_{s_0}(\tau,z)
       \prod_{j=1}^{R}\chi^{(N_j);\ell_j,s_j}_{m_j}(\tau,z),
\end{align*}
where $\chi^{\so(2)}_{s_0}(\tau,z)$ is the character of the affine
\so(2) constructed from the two free fermions of transverse directions of
spacetime, and each $\chi^{(N_j);\ell_j,s_j}_{m_1}(\tau,z)$ is the
contribution of the $j$th minimal model. $\lambda$ and $\mu$ are vectors
of labels defined as
\begin{align*}
 &\lambda:=(\ell_1,\dots,\ell_R),\\
 &\mu:=(s_0;s_1,\dots,s_R;m_1,\dots,m_R).
\end{align*}
Next we introduce an inner product between two $\mu$'s as
\begin{align*}
 &\mu\cdot\mu':=-\sum_{j=0}^{R}\frac{s_js'_j}{4}
                      +\sum_{j=1}^{R}\frac{m_jm'_j}{2N_j}.
\end{align*}
We also introduce $\beta$ vectors, which is the same type vector as
$\mu$, defined as
\begin{align*}
 &\beta_0:=(1;1,\dots,1;1,\dots,1),\\
 & \beta_j:=(2;0,\dots,0,\underset{\underset{S_j}{\wedge}}{2},
       0,\dots,0;0,\dots,0),\quad (j=1,\dots,R).
\end{align*}

By using these notations, the GSO condition (the total U(1) charge is to
be odd integer) and the condition of spin structure (all sub-theories are
to be in NS sector, or all to be in R sector) can be written as
\begin{align}
 & 2\beta_0\cdot \mu\in 2\Zb+1,\qquad \beta_j\cdot \mu\in\Zb,
    \quad j=1,\dots,R.
 \label{betaconstraint}
\end{align}
We can only use the modules satisfying the condition
(\ref{betaconstraint}) to construct the partition function. We call this
constraint (\ref{betaconstraint}) ``beta constraint''.

Let us define the ``orbit'' of the character $F^{\lambda}_{\mu}(\tau,z)$
for $\mu$'s satisfying the beta constraint (\ref{betaconstraint}) by the
equation
\begin{align*}
  &F^{\lambda}_{\mu}(\tau,z):=\sum_{b_0\in \Zb_{2K},b_j\in \Zb_2}
  (-1)^{s_0+b_0}\chi^{\lambda}_{\mu+\sum_{j=0}^{R}b_j\beta_j}(\tau,z),
\end{align*}
where $K:=\lcm(N_j,2)$.

By using this $F^{\lambda}_{\mu}$, the partition function can be written as
\begin{align*}
 Z(\tau,\bar\tau)=\frac{1}{\sqrt{\tau_2}|\eta|^2}\times
    \frac{1}{4^{R}2K}\sum_{\lambda,\mu}^{\rm beta}
      |F^{\lambda}_{\mu}(\tau,0)|^2,
\end{align*}
where the factor $\frac{1}{\sqrt{\tau_2}|\eta|^2}$ is the contribution
from the two free bosons of flat spacetime. The sum $\displaystyle
\sum_{\lambda,\mu}^{\rm beta}$ means that the labels satisfy the
beta constraint (\ref{betaconstraint}). Since there is spacetime
supersymmetry, this partition function should vanish. For the
partition function to vanish, the orbit $F^{\lambda}_{\mu}(\tau,0)$ 
should also vanish.

Because this Gepner model is a solvable realization of a Calabi-Yau
compactification, we expect the $F^{\lambda}_{\mu}$ is decomposed by
$\xi^{\su(3)}_{a}$'s , which mean, for some functions
$F^{\lambda}_{\mu,a}(\tau)$, we can write
\begin{align}
 F^{\lambda}_{\mu}(\tau,z)=\sum_{a\in\Zb_3}F^{\lambda}_{\mu,a}(\tau)
  \xi^{\su(3)}_{a}(\tau,z), \label{compactGepnerdecomposition}
\end{align}
where $z$ dependent $\xi^{\su(3)}_{a}(\tau,z)$ is defined simply as
\begin{align*}
 & \xi^{\su(3)}_{a}(\tau,z)=\frac{1}{\eta^2}\sum_{s\in\Zb_4}(-1)^{s}
     \Theta_{6+4a-3s,6}(\tau,z)
     \Theta_{s,2}(\tau,z).
\end{align*}
It reduces to the definition (\ref{su(3)holonomyidentity})
when we set $z=0$.

Now, let us show the decomposition (\ref{compactGepnerdecomposition})
and calculate the branching function $F^{\lambda}_{\mu,a}(\tau)$.
To do this, the product formula of the multiple theta functions is
useful. This formula can be written as
\begin{align}
  & \prod_{j=1}^{R}\Theta_{m_j,k_j}(\tau,z)
     =\sum_{r\in \Zb_k}B^{r}_{\{m_j\};\{k_j\}}(\tau)  
         \Theta_{\sum_j m_j+2r,k}(\tau,z),\nn\\
  & B^{r}_{\{m_j\};\{k_j\}}(\tau):=\sum_{\{n_j\}}\delta_{r-\sum_{j}k_j n_j,0}\ 
        q^{\sum_{j}k_j\left(n_j+\frac{m_j}{2k_j}\right)^2
              -\frac{1}{4k}\left[\sum_{j}(m_j+2k_jn_j)\right]^2},\nn\\
 &k:=\sum_{j=1}^{R}k_j,\qquad \sum_{\{n_j\}}:=\prod_{j=1}^{R}\sum_{n_j\in\Zb},
 \label{productformula}
\end{align}
where $q:=\e{\tau}$.
By using the formula (\ref{productformula}) and formulas in appendix
\ref{appendixA}, we can show the decomposition
(\ref{compactGepnerdecomposition})  is actually correct and
the branching function $F^{\lambda}_{\mu,a}(\tau)$'s are calculated
concretely
\begin{align*}
 &F^{\lambda}_{\mu,a}(\tau)=\eta(\tau)\sum_{b_j\in \Zb_2}
     \alpha^{\lambda}_{\mu+\sum_{j=1}^{R}b_j\beta_j,
       \ a-s_0-2\sum_jb_j+1}(\tau),\nn\\
 &\alpha^{\lambda}_{\mu,v}(\tau):=\frac12 \sum_{\{r_j\}}\sum_{v_0\in \Zb}
    \prod_{j=1}^{R}c^{(N_j-2)\ell_j}_{m_j-s_j-4r_j}(\tau)\sum_{\{p_j\}}
    \sum_{u\in \Zb_{3K}}
    B^{2Ku}_{\{2KQ_j(m_j,s_j)\};\{2KJ_j(N_j-2)\}}(\tau)\nn\\
    &\hspace{8cm} \times \delta_{2\sum_{j=1}^{R}Q_j(m_j,s_j)
                       +4u-s_0-2(6v_0+2v+1),0}\ ,\nn\\
 &\sum_{\{r_j\}}:=\prod_{j=1}^{R}\sum_{r_j\in\Zb_{N_j-2}} ,\qquad
    \sum_{\{p_j\}}:=\prod_{j=1}^{R}\sum_{p_j\in\Zb_{J_j}},
  \qquad J_j:=\frac{N_j}{K},\\
 &Q_j(m_j,s_j):=m_j/N_j-s_j/2+2r_j+2p_j(N_j-2).
\end{align*}
The detailed calculations are shown in appendix \ref{calculations}.

We have shown that the Gepner model includes the $\so(8)/\su(3)$ CFT in
the appropriate form. Especially, by using the character identity
$\xi^{\su(3)}_{a}(\tau,0)=0$ and the decomposition
(\ref{compactGepnerdecomposition}) we have shown 
the identity $F^{\lambda}_{\mu}(\tau,0)=0$, and the partition function
actually vanishes as expected.

We apply the same procedure to the models of
\cite{Yamaguchi:2000dz,Yamaguchi:2001rn} in the next section.

\section{Noncompact Gepner models}\label{noncompact}
Let us proceed to the noncompact Gepner models. Each of the noncompact
Gepner models includes the continuous part and discrete part due to
its noncompactness.

We consider each parts below, and show their partition functions
actually vanishes by using the character identities. For the continuous
part, it seems that they have twice as many supersymmetry charges as
expected from their holonomies. On the other hand, the discrete part
seems to have just the same amount of the supersymmetry as expected.

\subsection{Continuous part
 of the noncompact Calabi-Yau 4-fold compactifications}

We consider the continuous part of a noncompact Calabi-Yau 4-fold
compactification in this subsection. We show the partition function
includes the $\xi^{\su(3)}$ , which suggests this compactification has
twice as many supersymmetry charges as expected for the Calabi-Yau
4-fold.

First, we construct the partition function by the beta method, following
\cite{Yamaguchi:2000dz}.
In order to save the notations, we use the same notations of the beta
methods for the various compactifications.

The total theory is the direct product of the $\Ncal=2$ Liouville model, and
$R$ of minimal models with levels $(N_j-2),\ j=1,\dots,R$ . We define
the integer $K$ as $K=\lcm(N_j,2)$, and the integer $J$ as
$Q^2=\frac{2J}{K}$ , where $Q$ is the Liouville background charge. The
total character becomes
\begin{align*}
 &\chi^{\lambda}_{\mu}(\tau,z)=\chi^{\so(2)}_{s_0}(\tau,z)
      \frac{\Theta_{m_0,KJ}(\tau,2z/K)}{\eta(\tau)}
     \prod_{j=1}^{R}\chi^{(N_j);\ell_j,s_j}_{m_j}(\tau,z),\\
 &\lambda:=(\ell_1,\dots,\ell_R),\\
 &\mu:=(s_0;s_1,\dots,s_R;m_0;m_1,\dots,m_R),
\end{align*}
where $\chi^{\so(2)}_{s_0}$ is the character of 2 fermions in the $\Ncal=2$
Liouville model, and $\frac{\Theta_{m_0,KJ}(\tau,2z/K)}{\eta(\tau)}$ is the
character of the $S^1$ boson in $\Ncal=2$ Liouville.
We also define the inner product between two $\mu$ vectors as
\begin{align*}
 &\mu\cdot\mu'=-\frac{s_0s_0'}{4}-\sum_{j=1}^{R}\frac{s_js_j'}{4}
 -\frac{m_0m_0'}{2KJ}+\sum_{j=1}^{R}\frac{m_jm_j'}{2N_j}.
\end{align*}
We use the beta vectors to construct the modular invariant partition
function. The beta vectors are defined as
\begin{align*}
 &\beta_0:=(1;1,\dots,1;-J;1,\dots,1),\\
 &\beta_j:=(2;0,\dots,0,\underset{\underset{S_j}{\wedge}}{2},0,\dots,0;0;0,\dots,0),\quad (j=1,\dots,R).
\end{align*}
The GSO condition, and the spin structure condition can be written as
\begin{align}
 2\beta_0\cdot\mu\in 2\Zb+1,\qquad \beta_j\cdot\mu\in \Zb,\qquad (j=1,\dots,R).
 \label{beta4}
\end{align}
Let us call the above constraint ``the beta constraint''.
The orbits are defined for $\mu$ vectors which satisfy the beta
constraint (\ref{beta4}) as
\begin{align*}
 F^{\lambda}_{\mu}(\tau,z):=\sum_{b_0\in\Zb_{2K},b_j\in\Zb_{2}}
\chi^{\lambda}_{\mu+b_0\beta_0+\sum_{j=1}^{R}b_j\beta_j}(\tau,z)
(-1)^{s_0+b_0}.
\end{align*}
Using these notations, we can write down the modular invariant
partition function
\begin{align*}
 Z=\frac{1}{\sqrt{\tau_2}|\eta(\tau)|^2}\frac{1}{4^R 2K}
   \sum_{\lambda,\mu}^{\rm beta}|F^{\lambda}_{\mu}(\tau)|^{2},
\end{align*}
where the sum $\displaystyle \sum_{\lambda,\mu}^{\rm beta}$
 is taken under the beta constraint (\ref{beta4}).
This partition function can be checked to be actually modular
invariant\cite{Yamaguchi:2000dz}.

The spacetime supersymmetry implies $Z=0$ or equivalently
$F^{\lambda}_{\mu}(\tau,0)=0$ for $\lambda,\mu$ satisfying the beta
constraint. We prove this identity by showing the decomposition
\begin{align}
 F^{\lambda}_{\mu}(\tau,z)=\sum_{a\in\Zb_3}F^{\lambda}_{\mu,a}(\tau)
  \xi^{\su(3)}_{a}(\tau,z), \label{noncompactGepnerdecomposition4}
\end{align}
is possible for some branching function $F^{\lambda}_{\mu,a}(\tau)$.

We can show the decomposition is actually possible and obtain the
branching function concretely
\begin{align*}
 &F^{\lambda}_{\mu,a}(\tau)=\eta(\tau)\sum_{b_j\in \Zb_2}
     \alpha^{\lambda}_{\mu+\sum_{j=1}^{R}b_j\beta_j,\ 
   a-s_0-2\sum_{j=1}^{R}b_j+1}(\tau),\nn\\
 &\alpha^{\lambda}_{\mu,v}(\tau)
  :=\frac12\sum_{\{r_j\},\{p_j\}}\sum_{v_0\in\Zb}
   \left(\prod_{j=1}^{R}c^{(N_j-2);\ell_j}_{m_j-s_j-4r_j}(\tau)\right)
         \sum_{u\in \Zb_{3K}}B^{2Ku}_{\{2KQ_j\};\{\kappa_j\}}(\tau)
\delta_{2\sum_{j=0}^{R}Q_j+4u-s_0-2(6v_0+2v+1),0},
\end{align*}
where the $B^{2Ku}_{\{2KQ_j\};\{\kappa_j\}}(\tau)$'s are the branching
functions of theta functions defined by eq.(\ref{productformula}). We
also use the notations
\begin{align*}
 &\kappa_j:=2KJ_j(N_j-2),
 \qquad \kappa_0:=4KJ, \\
 &Q_j:=\frac{m_j}{N_j}-\frac{s_j}{2}+2r_j+2p_j(N_j-2),
 \qquad Q_0:=\frac{m_0}{K}+2Jp_0, \qquad (j=1,\dots,R),\\
 & \sum_{\{r_j\},\{p_j\}}=\sum_{p_0\in\Zb_2}\prod_{j=1}^{R}\left(
      \sum_{r_j\in\Zb_{N_j-2}}\sum_{p_j\in\Zb_{J_j}}\right),
 \qquad J_j=K/N_j.
\end{align*}
The decomposition (\ref{noncompactGepnerdecomposition4}) and
the identity $\xi^{\su(3)}(\tau)=0$ lead to
$F^{\lambda}_{\mu}(\tau)=0$ and $Z=0$.

Actually, the partition function includes $\xi^{\su(3)}$ as shown
eq.(\ref{noncompactGepnerdecomposition4}) and this fact suggests that
this compactification have the same amount of supersymmetry as
Calabi-Yau 3-fold compactifications, which is the twice as large amount
as the Calabi-Yau 4-fold compactifications.  This is an evidence that
the holographic dual of this singular compactification is a super {\em
conformal} field theory \cite{Giveon:1999zm}. The extra SUSY generators
are superconformal S generators in the superconformal field theory.

\subsection{Continuous part
 of the noncompact Calabi-Yau 3-fold compactifications}
We can perform the similar decomposition for the continuous part
 of a noncompact Calabi-Yau 3-fold compactification.

First, we construct the modular invariant partition functions by the
beta method.  The characters, $\mu$ vectors and the inner product
between them are defined as
\begin{align*}
& K:=\lcm(N_j,2),\qquad J:=\frac{K Q^2}{2},\\
&\chi^{\lambda}_{\mu}(\tau,z)=\chi^{\so(2)}_{s_{-1}}(\tau,z)
                \chi^{\so(2)}_{s_0}(\tau,z)
     \frac{\Theta_{m_0,KJ}(\tau,2z/K)}{\eta(\tau)}
     \prod_{j=1}^{R}\chi^{(N_j);\ell_j,s_j}_{m_j}(\tau,z),\\
 &\lambda:=(\ell_1,\dots,\ell_R),\\
 &\mu:=(s_{-1},s_0;s_1,\dots,s_R;m_0;m_1,\dots,m_R),\\
 &\mu\cdot\mu':=-\frac{s_{-1}s_{-1}'}{4}-\frac{s_0s_0'}{4}
 -\sum_{j=1}^{R}\frac{s_js_j'}{4}
 -\frac{m_0m_0'}{2KJ}+\sum_{j=1}^{R}\frac{m_jm_j'}{2N_j}.
\end{align*}
Here, the $\chi^{\so(2)}_{s_{-1}}(\tau,z)$ is the character of
two free fermions in transverse spacetime directions, and 
$\chi^{\so(2)}_{s_0}(\tau,z)$ is the character of the two free fermions
in $\Ncal=2$ Liouville theory.
The beta vectors are also defined as
\begin{align*}
& \beta_0:=(1,1;1,\dots,1;-J;1,\dots,1),\\
& \beta_{-1}:=(2,2;0,\dots,0;0;0,\dots,0),\\
&\beta_j:=(0,2;0,\dots,0,\underset{\underset{S_j}{\wedge}}{2},0,\dots,0;0;0,\dots,0),\quad (j=-1,1,\dots,R).
\end{align*}
The beta constraint can be written as
\begin{align*}
 2\beta_0\cdot\mu\in 2\Zb+1,\qquad \beta_j\cdot\mu\in \Zb,\qquad
    (j=-1,1,\dots,R).
\end{align*}
The orbits and the partition functions can be written as
\begin{align*}
 & F^{\lambda}_{\mu}(\tau,z):=
      \sum_{b_0\in\Zb_{2K},b_{-1}\in\Zb_2,b_j\in\Zb_{2}}
    \chi^{\lambda}_{\mu+b_0\beta_0+b_{-1}\beta_{-1}+\sum_{j=1}^{R}b_j\beta_j}
      (\tau,z)(-1)^{s_0+b_0},\\
 & Z(\tau,\bar \tau)
    =\left(\sqrt{\tau_2}|\eta(\tau)|^2\right)^{-3}\frac{1}{4^R 4K}
   \sum_{\lambda,\mu}^{\rm beta}|F^{\lambda}_{\mu}(\tau,0)|^{2},
\end{align*}
where $\left(\sqrt{\tau_2}|\eta(\tau)|^2\right)^{-3}$ is the
contribution from the two free bosons in the transverse spacetime
directions and the linear dilaton in the $\Ncal=2$ Liouville theory.

The spacetime supersymmetry implies $Z=0$ and
$F^{\lambda}_{\mu}(\tau,0)=0$. We prove this fact by showing the
decomposition
\begin{align*}
 F^{\lambda}_{\mu}(\tau,z)=\sum_{a\in\Zb_2}F^{\lambda}_{\mu,a}(\tau)
 \xi^{\su(2)}_{a}(\tau,z),
\end{align*}
for some branching function $F^{\lambda}_{\mu,a}(\tau)$.
Here we use the $z$ dependent $\xi^{\su(2)}(\tau,z)$ defined as
\begin{align*}
 \xi^{\su(2)}_{a}(\tau,z):=\sum_{s\in\Zb_4}
     \chi^{\so(2)}_{s}(\tau,z)\chi^{\so(2)}_{2a+s}(\tau,z)
    \frac{\Theta_{a+s+1,1}(\tau,2z)}{\eta(\tau)}(-1)^{s}.
\end{align*}
These $\xi^{\su(2)}_{a}(\tau,z)$ reduce to the functions defined in eq.
(\ref{su(2)holonomyidentity}) when we set $z=0$.

The branching functions are obtained as follows.
\begin{align*}
 &F^{\lambda}_{\mu,a}(\tau)=\sum_{\{b_j\}}
  \sum_{v\in\Zb_2}\alpha^{\lambda}_{\mu+\sum_{j=1}^{R}b_j\beta_j,v}(\tau)
     \deltam{2a-(s_{-1}-s_0-2\sum_{j=1}^{R}b_{j})}{4},\\
  &\alpha^{\lambda}_{\mu,v}(\tau)
  :=\frac12\sum_{\{r_j\},\{p_j\}}\sum_{v_0\in\Zb}
   \left(\prod_{j=1}^{R}c^{(N_j-2);\ell_j}_{m_j-s_j-4r_j}(\tau)\right)
         \sum_{u\in \Zb_{2K}}B^{2Ku}_{\{2KQ_j\};\{\kappa_j\}}(\tau)\\
 &\hspace{7cm}\times\delta_{2\sum_{j=0}^{R}Q_j+4u-s_{-1}-s_0-2(4v_0+2v+1),0},
\end{align*}
where the $B^{2Ku}_{\{2KQ_j\};\{\kappa_j\}}(\tau)$'s are the branching
functions of theta functions defined by eq.(\ref{productformula}). We
also use the notations
\begin{align*}
 &\kappa_j:=2KJ_j(N_j-2),
 \qquad \kappa_0:=4KJ, \\
 &Q_j:=\frac{m_j}{N_j}-\frac{s_j}{2}+2r_j+2p_j(N_j-2),
 \qquad Q_0:=\frac{m_0}{K}+2Jp_0, \qquad (j=1,\dots,R).\\
 & \sum_{\{r_j\},\{p_j\}}=\sum_{p_0\in\Zb_2}\prod_{j=1}^{R}\left(
      \sum_{r_j\in\Zb_{N_j-2}}\sum_{p_j\in\Zb_{J_j}}\right),
 \qquad J_j=K/N_j.
\end{align*}
The decomposition (\ref{noncompactGepnerdecomposition4}) and
the identity $\xi^{\su(2)}(\tau)=0$ lead to
$F^{\lambda}_{\mu}(\tau)=0$ and $Z=0$.

In the case of the continuous part in a noncompact Calabi-Yau 3-fold
compactification, the result suggests that this system has the same
amount of supersymmetry as the Calabi-Yau 2-fold compactifications.
The extra supersymmetry charges seem to correspond to superconformal S
generators in the dual superconformal field theory.

\subsection{Comments on the continuous part
 of the noncompact Calabi-Yau 2-fold compactifications and $G_2$
 compactifications}

As for the continuous part of a noncompact Calabi-Yau 2-fold
compactification (an ALE compactification), it is shown in
\cite{Eguchi:2000tc,Naka:2000uy} that the partition functions of all
models of this type includes the $\xi^{\{1\}}$ and vanish due to the
Jacobi's abstruse identity itself. This result shows that the models of
this type have the same amount of the supersymmetry as the flat
10-dimensional model, that is, twice as many supersymmetry as the ALE
compactification.  This fact is consistent with the conjecture that an
actually singular ALE compactification is holographically dual to a 6
dimensional superconformal field theory.

We also comment about the noncompact $G_2$ holonomy models obtained in
\cite{Eguchi:2001xa}. The partition functions of these models include
the $\xi^{\su(3)}$. This fact suggests that these models have twice as
many supercharges as expected for a $G_2$ compactification. For this
reason, these actually singular models should be also holographically dual to
3-dimensional superconformal theories.

\subsection{The discrete part of
  the ALE compactifications}

In the noncompact theory, besides the continuous part of the spectrum
which we treat in previous subsections, there is the discrete part of the
spectrum.  In this subsection, we consider the discrete part of the
spectrum of ALE models \cite{Yamaguchi:2001rn}.

We use the character of the {\slr/U(1)} Kazama-Suzuki model.
The character of the discrete series of the {\slr/U(1)} Kazama-Suzuki
model summed through the {\slr} spectral flow orbit can be written as
\begin{align*}
  &\chic^{(N)\ell,s}_{m}=\sum_{r\in \Zb_{N-2}}
    \cc^{(N+2);\ell}_{m-s-4r}(\tau)
     \Theta_{-2m-Ns+4Nr,2N(N+2)}(\tau,z/N),
\end{align*}
where $\cc^{(k);\ell}_{m}(\tau)$ is the ``{\slr} string function''.  The
explicit form of this string function is written in
\cite{Yamaguchi:2001rn}, but we do not need it here.

Again, we construct the partition function by the beta method.
We use the following notations
\begin{align*}
 & K:=\lcm(N,2), \qquad J:=K/N, \\
 &\chi^{\lambda}_{\mu}=\chi^{\so(2)}_{s_{-1}}(\tau,z)
                \chi^{\so(2)}_{s_0}(\tau,z)
             \chic^{(N)\ell_1,s_1}_{m_1}(\tau,z)
             \chi^{(N)\ell_2,s_2}_{m_2}(\tau,z),\\
 &\lambda:=(\ell_1,\ell_2),\qquad \mu:=(s_{-1},s_0;s_1,s_2;m_1,m_2),\\
 &\mu\cdot\mu'=-\frac{s_{-1}s_{-1}'}{4}-\frac{s_0s_0'}{4}
         -\frac{s_{1}s_{1}'}{4}-\frac{s_2s_2'}{4}
         -\frac{m_1m_1'}{2N} +\frac{m_2m_2'}{2N},\\
 &\beta_0:=(1,1;1,1;1,1),\qquad \beta_{-1}:=(2,2;0,0;0,0),\\
 &\beta_{1}:=(0,2;2,0;0,0),\qquad \beta_{2}:=(0,2;0,2;0,0),\\
\end{align*}
The orbit and the partition function can be written as
\begin{align*}
 &F^{\lambda}_{\mu}:=\sum_{b_0\in\Zb_{2K},\ b_{-1},b_{1},b_{2}\in\Zb_{2}}
\chi^{\lambda}_{\mu+b_0\beta_0+b_{-1}\beta_{-1}+
         b_{1}\beta_{1}+b_{2}\beta_{2}}(\tau,z)(-1)^{s_0+b_0},\\
 &Z=\left(\sqrt{\tau_2}|\eta(\tau)|^2\right)^{-2}\frac{1}{2^2 2^3 2K}
   \sum_{\lambda,\mu}^{\rm beta}|F^{\lambda}_{\mu}|^{2}.
\end{align*}

We can again show $Z=0,\ F^{\lambda}_{\mu}=0$ by showing the
decomposition
\begin{align*}
 F^{\lambda}_{\mu}(\tau,z)=\sum_{a\in\Zb_2}F^{\lambda}_{\mu,a}(\tau)
 \xi^{\su(2)}_{a}(\tau,z).
\end{align*}
The $F^{\lambda}_{\mu,a}(\tau)$ can be written as
\begin{align*}
 &F^{\lambda}_{\mu,a}(\tau)=\sum_{\{b_j\}}
  \sum_{v\in\Zb_2}\alpha^{\lambda}_{\mu+\sum_{j=1}^{R}b_j\beta_j,v}(\tau)
     \deltam{2a-(s_{-1}-s_0-2\sum_{j=1}^{R}b_{j})}{4},\\
  &\alpha^{\lambda}_{\mu,v}(\tau)
  :=\frac12\sum_{\{r_j\},\{p_j\}}\sum_{v_0\in\Zb} 
 \left(\cc^{\ell_1}_{m_1-s_1-4r_1}(\tau)
        c^{\ell_2}_{m_2-s_2-4r_2}(\tau)\right) \\
 &\hspace{5cm} \times 
   \sum_{u\in \Zb_{2K}}B^{2Ku}_{\{2KQ_j\};\{\kappa_j\}}(\tau)
    \delta_{2(Q_1+Q_2)+4u-s_{-1}-s_0-2(4v_0+2v+1),0}.
\end{align*}
Here we use the notations
\begin{align*}
 &\kappa_1:=2KJ(N+2),\qquad
 \kappa_2:=2KJ(N-2),\\
 &Q_1:=-\frac{m_1}{N}-\frac{s_1}{2}+2r_1+2p_1(N+2),\qquad
 Q_2:=\frac{m_2}{N}-\frac{s_2}{2}+2r_2+2p_2(N-2).
\end{align*}

In contrast to the continuous part,
the discrete part seems to have just the same amount of the
supersymmetry as expected for an ALE compactification.
This is because the discrete part exists only in the theory of 
deformed singularity, and the holographic dual theory is
relevantly perturbed and does not have conformal symmetry
\cite{Giveon:1999px,Giveon:1999tq}.

\section{Conclusion}\label{conclusion}

In this paper, we show the partition functions of several supersymmetric
string models include the coset characters with their vanishing forms.
We have shown that all the partition functions of the supersymmetric
models treated here {\em do vanish}.

The compact Gepner models and the discrete part of the noncompact Gepner
models include the appropriate characters with their supersymmetry. But
the partition functions of continuous part of noncompact Gepner models
look as if they have twice as many supersymmetry as expected.  We claim
that this fact shows the holographic dual of an actually singular
compactification is a super {\em conformal} field theory, and the extra
supercharges correspond to superconformal S generators in the
holographic dual superconformal field theory.

On the other hand, the discrete part exist only in the deformed
singularity, and the holographic dual theory is no longer conformally
invariant and has renormalization group flow.  This is why the partition
function of the discrete part includes just the same amount of the
supersymmetry as expected.

\subsection*{Acknowledgement}
I would like to thank Katsuyuki Sugiyama and Tsuneo Uematsu for useful
discussions and encouragement.  I would also like to thank the
organizers of the Summer Institute 2001 at Yamanashi, Japan, 6-20
August, 2001, where a part of this work is done.

The work of the author is supported in part by the JSPS Research
Fellowships for Young Scientists.
\newpage
\appendix
\section{Theta functions and characters}\label{appendixA}
In this appendix A, we collect several notations and summarize
properties of theta functions.  We use the following notations in this
paper;
\begin{eqnarray*}
 &&\e{x}:=\exp(2\pi i x),
 \qquad \deltam{m}{N}:=
\begin{cases}
 1 & (m\equiv 0 \mod N),\\
 0 & ({\rm others}),
\end{cases}\\
\end{eqnarray*}
where $m$ and $N$ are integers. 
For a function $f(\tau,z)$, we sometimes use abbreviated notation
\begin{align*}
 f:=f(\tau):=f(\tau,z=0).
\end{align*}

\subsection{Theta functions}
A set of SU(2) classical theta functions are defined as
\begin{eqnarray*}
&& \Th_{m,k}(\tau,z)=\sum_{n\in \Zb}q^{k\left(n+\frac{m}{2k}\right)^2}
 y^{k\left(n+\frac{m}{2k}\right)},
\end{eqnarray*}
with $q:=\e{\tau},y:=\e{z}$. We often uses the formulas for integer $m,k,p$
\begin{align}
 & \Theta_{m/p,k/p}(\tau,z)=\sum_{t\in \Zb_{p}}\Theta_{m+2kt,pk}(\tau,z/p).
\label{fractionaltheta}
\end{align}
We also use the following product formula of the theta functions
\begin{align}
 &\Theta_{m_1,k_1}(\tau,z_1)\Theta_{m_2,k_2}(\tau,z_2)
=\sum_{r\in \Zb_{k_1+k_2}}
\Theta_{m_2k_1-m_1k_2+2k_1k_2r,k_1k_2(k_1+k_2)}(\tau,u)
\Theta_{m_1+m_2+2k_2r,k_1+k_2}(\tau,v),\nn\\
 &\qquad u=\frac{z_2-z_1}{k_1+k_2},\qquad v=\frac{k_1z_1+k_2z_2}{k_1+k_2}.
\label{2product-formula}
\end{align}
The Jacobi's theta functions are also
defined in our convention
\begin{eqnarray*}
&& \theta_{1}(\tau,z):=i\sum_{n\in \Zb}(-1)^n q^{\frac12\left(n-\frac{1}{2}\right)^2}
 y^{\left(n-\frac{1}{2}\right)},
 \theta_{2}(\tau,z):=\sum_{n\in \Zb} q^{\frac12\left(n-\frac{1}{2}\right)^2}
 y^{\left(n-\frac{1}{2}\right)},\\
&& \theta_{3}(\tau,z):=\sum_{n\in \Zb} q^{\frac12n^2}y^{n},\hspace{2.7cm}
 \theta_{4}(\tau,z):=\sum_{n\in \Zb}(-1)^n q^{\frac12n^2}y^{n}.
\end{eqnarray*}
The above two kinds of theta functions are related through a set of
linear transformations
\begin{eqnarray*}
 && 2\Th_{0,2}(\tau,z)=\theta_3(\tau,z)+\theta_4 (\tau,z),
\quad 2\Th_{1,2}(\tau,z)=\theta_2(\tau,z)+i\theta_1(\tau,z),\\ 
 && 2\Th_{2,2}(\tau,z)=\theta_3(\tau,z)-\theta_4(\tau,z) ,\quad
   2\Th_{3,2}(\tau,z)=\theta_2(\tau,z)-i\theta_1(\tau,z).
\end{eqnarray*}
The Dedekind $\eta$ function is represented as an infinite product
\begin{eqnarray*}
 \eta(\tau):=q^{\frac1{24}}\prod_{n=1}^{\infty}(1-q^n).
\end{eqnarray*}

\subsection{$\Ncal=2$ minimal models}
The unitary minimal models of the $\Ncal=2$ superconformal algebra
is labeled by an integer $k=1,2,3,\dots$ . Instead of $k$ we mainly
use the ``dual Coxeter number'' of ADE classification $N=k+2$.
The Verma module of the level $(N-2)$ minimal model is labeled by
a set of three integers $(\ell,m,s)$
\begin{align*}
 \ell=0,1,\dots,N-2,\qquad
 m\in \Zb_{2N},\qquad
 s\in \Zb_{4},\\
 \ell+m+s\equiv 0 \mod 2,\qquad (\ell,m,s)\cong (N-2-\ell,m+N,s+2).
\end{align*}
We introduce a character $\chi_m^{(N);\ell,s}(\tau,z)$ of a Verma module
$(\ell,m,s)$ in the level $(N-2)$ minimal model.  The form of
the character $\chi_m^{(N);\ell,s}(\tau,z)$ can be written as
\begin{align*}
 \chi_m^{(N);\ell,s}(\tau,z)=\sum_{r\in \Zb_{N-2}}c^{(N-2);\ell}_{m-s+4r}
 \Theta_{2m+N(-s+4r),2N(N-2)}(\tau,z/N),
\end{align*}
where $c^{(N-2);\ell}_{m}$'s are the string functions of the level 
$(N-2)$ affine SU(2),
defined by the branching relation
\begin{align*}
 \chi^{SU(2),(N-2)}_{\ell}(\tau,z)
    =\sum_{m\in\Zb_{N-2}}c^{(N);\ell}_{m}(\tau)
    \Theta_{m,N-2}(\tau,z).
\end{align*}
In this equation, $\chi^{SU(2),(k)}_{\ell}(\tau,z)$'s are the level $k$
affine SU(2) characters expressed by the Weyl-Kac formula
\begin{align*}
 \chi^{SU(2),(k)}_{\ell}(\tau,z)=
     \frac{\Theta_{\ell+1,k+2}(\tau,z)-\Theta_{-\ell-1,k+2}(\tau,z)}%
       {\Theta_{1,2}(\tau,z)-\Theta_{-1,2}(\tau,z)}.
\end{align*}

\subsection{Virasoro minimal models} \label{appendix-minimal}
The unitary minimal models are labeled by an integer $m$
($m=3,4,5,\dots$). 
Its central
charge is given by a formula
\begin{align*}
 c=1-\frac{6}{m(m+1)}.
\end{align*}
The Verma modules of each minimal model is classified by integers $r,s$ in the
regions
\begin{align*}
 r=1,2, \dots,m-1,\qquad s=1,2, \dots,m,\qquad \mbox{with}\,\,ms<(m+1)r.
\end{align*}
The conformal dimension of
the primary field is specified by the set $(r,s)$ and is evaluated as
\begin{align*}
 h_{r,s}=\frac{\{(m+1)r-ms\}^2-1}{4m(m+1)}.
\end{align*}
The characters of these minimal models can be expressed for the 
primary field labelled by $(r,s)$
\begin{align*}
 \chi^{(m)}_{r,s}=\frac{1}{\eta(\tau)}
\{\Theta_{(m+1)r-ms,m(m+1)}(\tau)-\Theta_{(m+1)r+ms,m(m+1)}(\tau)\}.
\end{align*}
We use $m=3,4,5$ minimal models in this paper. The details of properties
of these models are listed in the following table:
\begin{itemize}
 \item \underline{Ising model\,\,} \qquad ($c=\frac12$)
 \begin{align*}
  h_{1,1}=0,\ h_{2,1}=\frac12,\ h_{1,2}=\frac1{16}.
 \end{align*}
 We write the Virasoro characters for this model as $\chi^{\isi}_{h_{r,s}}$.
 \item \underline{Tricritical Ising model\,\,} \qquad ($c=\frac{7}{10}$)
 \begin{align*}
  h_{1,1}=0,\ h_{2,1}=\frac{7}{16},\ h_{1,2}=\frac 1{10},\ 
h_{1,3}=\frac 35, \ h_{2,2}=\frac 3{80},\ h_{3,1}=\frac 32.
 \end{align*}
 We write the Virasoro characters of this model as $\chi^{\tri}_{h_{r,s}}$.
 \item \underline{3-state Potts model\,\,} \qquad ($c=\frac{4}{5}$)
 \begin{align*}
  h_{1,1}=0,\ h_{2,1}=\frac{2}{5},\ h_{3,1}=\frac 75,\ 
h_{1,3}=\frac 23, \ h_{4,1}=3,\ h_{2,3}=\frac 1{15}.
 \end{align*}
The notation $\chi^{\pot}_{h_{r,s}}$ is used for Virasoro characters 
for this Potts model. But we 
mainly use $W_3$ characters constructed from those of the 
Potts model
\begin{align*}
& C^{\pot}_{0}=\chi^{\pot}_{0}+\chi^{\pot}_{3},\quad
& C^{\pot}_{2/5}=\chi^{\pot}_{2/5}+\chi^{\pot}_{7/5}, \\
& C^{\pot}_{2/3}=\chi^{\pot}_{2/3},\quad
& C^{\pot}_{1/15}=\chi^{\pot}_{1/15}.\quad
\end{align*}
The standard modular invariant partition function of the 3-state Potts model
can be described by using these $W_3$ characters $C^{\pot}$'s 
\begin{align*}
 Z=|C^{\pot}_{0}|^2+|C^{\pot}_{2/5}|^2+2|C^{\pot}_{2/3}|^2
+2|C^{\pot}_{1/15}|^2.
\end{align*}
\end{itemize}
\subsection{Level 1 $\so(2r)$ WZW models}
\label{appendix-WZW}
We denote the character of  the level $1$ affine $\so(2r)$ 
as $\chi^{\so(2r)}_{s}(\tau),\  s=0,1,2,3$. The explicit forms are given as
\begin{align*}
& \chi^{\so(2r)}_{0}(\tau)=\frac{1}{2\eta(\tau)^r}\left(
\theta_3(\tau)^r+\theta_4(\tau)^r\right),\qquad
& \chi^{\so(2r)}_{2}(\tau)=\frac{1}{2\eta(\tau)^r}\left(
\theta_3(\tau)^r-\theta_4(\tau)^r\right),\\
& \chi^{\so(2r)}_{1}(\tau)=\chi^{\so(2r)}_{3}(\tau)
 =\frac{1}{2\eta(\tau)^r}\theta_2(\tau)^r.\\
\end{align*}
We sometimes use the $z$ dependent character of the level $1$ affine $\so(2)$
\begin{align*}
 \chi^{\so(2)}_{s}(\tau,z)=\frac{\Theta_{s,2}(\tau,z)}{\eta(\tau)}.
\end{align*}

\section{Detailed calculations}\label{calculations}
We show here the detailed calculations of the decomposition, in the case
of the compact Gepner model treated in section
\ref{compact}. We use here the notation of the beta method in section
\ref{compact}. The other models treated in this paper can be calculated
in almost the same manner.

First we show the product formula of multiple theta functions
\begin{align*}
 \prod_{j=1}^{R}\Theta_{m_j,k_j}(\tau,z)&=\sum_{\{n_j\}}
   q^{\sum_{j}k_j\left(n_j+\frac{m_j}{2k_j}\right)^2}
   y^{\sum_{j}k_j\left(n_j+\frac{m_j}{2k_j}\right)}
 &=\sum_{\{n_j\}}
   q^{\sum_{j}k_j\left(n_j+\frac{m_j}{2k_j}\right)^2}
   y^{\sum_{j}k_j n_j+\frac12 \sum_{j}m_j}.
\end{align*}
If we insert the identity
\begin{align*}
  1=\sum_{n\in \Zb,\ r\in \Zb_k}\delta_{kn+r-\sum_{j}k_j n_j,0},\qquad
 k:=\sum_{j=1}^{R}k_j,
\end{align*}
then the product becomes
\begin{align*}
  \prod_{j=1}^{R}\Theta_{m_j,k_j}(\tau,z)
     &=\sum_{r\in \Zb_k}\sum_{n\in \Zb}\sum_{\{n_j\}}
        \delta_{kn+r-\sum_{j}k_j n_j,0}\;
        q^{\sum_{j}k_j\left(n_j+\frac{m_j}{2k_j}\right)^2}
        y^{k\left(n+\frac{\sum_{j}m_j+2r}{2k}\right)}\\
     &=\sum_{r\in \Zb_k}\sum_{n\in \Zb}B^{r}_{\{m_j\};\{k_j\}}(\tau)  
         q^{k\left(n+\frac{\sum_{j}m_j+2r}{2k}\right)^2}
         y^{k\left(n+\frac{\sum_{j}m_j+2r}{2k}\right)},\\
 B^{r}_{\{m_j\};\{k_j\}}(\tau)
    &:=
        \sum_{\{n_j\}}\delta_{kn+r-\sum_{j}k_j n_j,0}\ 
        q^{\sum_{j}k_j\left(n_j+\frac{m_j}{2k_j}\right)^2
              -k\left(n+\frac{\sum_{j}m_j+2r}{2k}\right)^2}.
\end{align*}
If we shift $n_j\to n_j+n$, $B^{r}_{\{m_j\};\{k_j\}}(\tau)$
 can be rewritten as
\begin{align*}
  B^{r}_{\{m_j\};\{k_j\}}(\tau)
   &=
        \sum_{\{n_j\}}\delta_{r-\sum_{j}k_j n_j,0}\ 
        q^{\sum_{j}k_j\left(n_j+n+\frac{m_j}{2k_j}\right)^2
              -k\left(n+\frac{\sum_{j}m_j+2r}{2k}\right)^2} \\
  &=\sum_{\{n_j\}}\delta_{r-\sum_{j}k_j n_j,0}\ 
        q^{\sum_{j}k_j\left(n_j+\frac{m_j}{2k_j}\right)^2
              -\frac{1}{4k}\left[\sum_{j}(m_j+2k_jn_j)\right]^2},
\end{align*}
This $B^{r}_{\{m_j\};\{k_j\}}(\tau)$ is actually independent of $n$.
Then, the product becomes
\begin{align*}
   \prod_{j=1}^{R}\Theta_{m_j,k_j}(\tau,z)
     =\sum_{r\in \Zb_k}B^{r}_{\{m_j\};\{k_j\}}(\tau)  
         \Theta_{\sum_j m_j+2r,k}(\tau,z).
\end{align*}
This is the product formula (\ref{productformula}).
If $K$ is a common divisor of $\{k_j\}$,
\begin{align*}
 B^{r}_{\{m_j\};\{k_j\}}(\tau)=0, \qquad \text{if } r\not \equiv 0 \mod K.
\end{align*}
Then the product formula becomes
\begin{align*}
    \prod_{j=1}^{R}\Theta_{m_j,k_j}(\tau,z)
     =\sum_{r\in \Zb_{k/K}}B^{Kr}_{\{m_j\};\{k_j\}}(\tau)  
         \Theta_{\sum_j m_j+2Kr,k}(\tau,z).
\end{align*}
Note that for any $a\in \Rb$ the relation
\begin{align*}
 B^{r}_{\{m_j+k_j a\};\{k_j\}}=B^{r}_{\{m_j\};\{k_j\}},
\end{align*}
is satisfied.

Let us proceed to the branching function of the orbit.
The character of a $\Ncal=2$ minimal model can be written as
\begin{align*}
 \chi^{(N_j);\ell_j,s_j}_{m}(\tau,z)&=\sum_{r_j\in \Zb_{N_j-2}}
   c^{(N_j-2);\ell_j}_{m_j-s_j-4r_j}(\tau)
      \Theta_{2m_j-N_js_j+4N_jr_j,2N_j(N_j-2)}
         (\tau,z/N_j)\\
  &=\sum_{r\in \Zb_{N-2}}
   c^{(N_j-2);\ell_j}_{m_j-s_j-4r_j}(\tau)\sum_{p_j\in \Zb_{J_j}}
     \Theta_{2K[m_j/N_j-s_j/2+2r_j+2p_j(N_j-2)],2KJ_j(N_j-2)}(\tau,z/K).
\end{align*}
Here, we use the formula (\ref{fractionaltheta})
The total character becomes
\begin{align*}
 \chi^{\lambda}_{\mu}(\tau,z)
   &=\chi^{\so(2)}_{s_0}(\tau,z)\prod_{j=1}^{R}
                 \chi^{(N_j);\ell_j,s_j}_{m_j}(\tau,z)\\
   &=\chi^{\so(2)}_{s_0}(\tau,z)\sum_{\{r_j\}}
    \prod_{j=1}^{R}c^{(N_j-2);\ell_j}_{m_j-s_j-4r_j}(\tau)\sum_{\{p_j\}}
    \prod_{j=1}^{R}\Theta_{2K[m_j/N_j-s_j/2+2r_j+2p_j(N_j-2)],2KJ_j(N_j-2)}
    (\tau,z/K)
\end{align*}
If we use the product formula of multiple theta functions and
$\sum_{j=1}^{R}2KJ_j(N_j-2)=6K^2$, the total character becomes
\begin{align*}
  \chi^{\lambda}_{\mu}(\tau,z)
   =\chi^{\so(2)}_{s_0}(\tau,z)\sum_{\{r_j\}}
    \prod_{j=1}^{R}c^{(N_j-2);\ell_j}_{m_j-s_j-4r_j}(\tau)\sum_{\{p_j\}}
    \sum_{u\in \Zb_{3K}}
    B^{2Ku}_{\{2K[m_j/N_j-s_j/2+2r_j+2p_j(N_j-2)]\};\{2KJ_j(N_j-2)\}}(\tau)\\
 \times
       \Theta_{2K\sum_j[m_j/N_j-s_j/2+2r_j+2p_j(N_j-2)]+4Ku,6K^2}
    (\tau,z/K)
\end{align*}
Let us define $Q_j(m_j,s_j)=m_j/N_j-s_j/2+2r_j+2p_j(N_j-2)$, then
this is the charge contribution of the $j$th minimal model modulo $2$.
We can calculate the sum
\begin{align*}
 \sum_{a_0\in \Zb_{K/2}}\chi^{\lambda}_{\mu+4a_0\beta_0}
  &=\frac12 \sum_{a_0\in \Zb_{K}}\chi^{\lambda}_{\mu+4a_0\beta_0}\\
  &=\frac12 \chi^{\so(2)}_{s_0}(\tau,z)\sum_{\{r_j\}}
    \prod_{j=1}^{R}c^{(N_j-2);\ell_j}_{m_j-s_j-4r_j}(\tau)\sum_{\{p_j\}}
    \sum_{u\in \Zb_{3K}}
    B^{2Ku}_{\{2KQ_j(m_j,s_j)\};\{2KJ_j(N_j-2)\}}(\tau)
    \\&\hspace{3cm}\times\sum_{a_0\in \Zb_K}
      \Theta_{2K\sum_jQ_j(m_j,s_j)-12 Ka+4Ku,6K^2}
       (\tau,z/K)\\
  &=\frac12 \chi^{\so(2)}_{s_0}(\tau,z)\sum_{\{r_j\}}
    \prod_{j=1}^{R}c^{(N_j-2);\ell_j}_{m_j-s_j-4r_j}(\tau)\sum_{\{p_j\}}
    \sum_{u\in \Zb_{3K}}
    B^{2Ku}_{\{2KQ_j(m_j,s_j)\}}(\tau)
       \\&\hspace{3cm}\times\Theta_{2\sum_jQ_j(m_j,s_j)+4u,6}
       (\tau,z),
\end{align*}
where we use the formula (\ref{fractionaltheta}). Due to the
GSO condition, the sum of the charge
$\sum_jQ_j(m_j,s_j)-s_0/2=(\text{odd})$, and the following identity is
satisfied.
\begin{align*}
  1
  =\sum_{v_0\in \Zb,\ v\in \Zb_{3}}
      \delta_{2\sum_jQ_j(m_j,s_j)+4u-s_0-2(6v_0+2v+1),0}.
\end{align*}
Using these formulae, we obtain the sum
\begin{align*}
   &\sum_{a_0\in \Zb_{K/2}}\chi^{\lambda}_{\mu+4a_0\beta_0}
   =\sum_{v\in \Zb_3}
   \alpha^{\lambda}_{\mu,v}(\tau)
    \chi^{\so(2)}_{s_0}\Theta_{s_0+4v+2,6}(\tau,z),\\
&\alpha^{\lambda}_{\mu,v}(\tau)
:=\frac12 \sum_{\{r_j\}}\sum_{v_0\in \Zb}
    \prod_{j=1}^{R}c^{(N_j-2);\ell_j}_{m_j-s_j-4r_j}(\tau)\sum_{\{p_j\}}
    \sum_{u\in \Zb_{3K}}
    B^{2Ku}_{\{2KQ_j(m_j,s_j)\};\{2KJ_j(N_j-2)\}}(\tau)\\ &\hspace{4cm}
  \times  \delta_{2\sum_jQ_j(m_j,s_j)+4u-s_0-2(6v_0+2v+1),0}.
\end{align*}
This $\alpha^{\lambda}_{\mu,v}(\tau)$ satisfies the relation
\begin{align*}
 &\alpha^{\lambda}_{\mu+c_0\beta_0,v}=\alpha^{\lambda}_{\mu,v+c_0},
\qquad \text{for } \ c_0\in\Zb.
\end{align*}
By using this relation, we obtain the sum
\begin{align*}
\sum_{c_0\in \Zb_4}\sum_{a_0\in \Zb_{K/2}}
   (-1)^{c_0+s_0}
  \chi^{\lambda}_{\mu+c_0\beta_0+4a_0\beta_0}(\tau,z)
=\sum_{a\in \Zb_3} \alpha^{\lambda}_{\mu,a-s_0+1}\eta(\tau)
\xi^{\su(3)}_{a}(\tau,z),
\end{align*}
and the decomposition of orbit becomes
\begin{align*}
 &F_{\mu}^{\lambda}(\tau,z)
   =\sum_{a\in\Zb_3}F^{\lambda}_{\mu,a}(\tau)\xi^{\su(3)}_a(\tau,z),\\
 &F^{\lambda}_{\mu,a}(\tau):=\eta(\tau)\sum_{b_j\in \Zb_2}
     \alpha^{\lambda}_{\mu+\sum_jb_j\beta_j,\ a-s_0-2\sum_jb_j+1}(\tau).
\end{align*}

Also in the case of the noncompact Gepner models, we can decompose
the orbit in almost the same manner.

\newpage
\providecommand{\href}[2]{#2}\begingroup\raggedright\endgroup

\end{document}